\documentclass[prb,twocolumn,showpacs,floatfix]{revtex4}
\usepackage{graphicx}
\usepackage{dcolumn}
\usepackage{bm}

\begin{document}

\preprint{}

\title{Electric field-induced spin textures in a superlattice with Rashba and Dresselhaus
spin-orbit coupling}

\author{D.V. Khomitsky}
 \email{khomitsky@phys.unn.ru}
 \affiliation{Department of Physics, University of Nizhny Novgorod,
              23 Gagarin Avenue, 603950 Nizhny Novgorod, Russian Federation}

\date{\today}

\begin{abstract}

The DC charge current, field-induced spin polarization and spin textures are studied in the 1D gated superlattice
with both fixed and varying Rashba and Dresselhaus contributions to spin-orbit coupling. It is found that
a spin component with zero mean value can demonstrate non-vanishing field-induced spin texture
in a superlattice cell which can be probed experimentally, with the highest amplitudes achievable in
an interval of comparable Rashba and Dresselhaus terms. The consideration of the finite parameters for collision rate
and temperature is found to be non-destructive for the calculated current and spin characteristics depending on all
states below the Fermi level.

\end{abstract}


\maketitle

\section{Introduction}

The spin polarization of charge carriers in nanostructures is an important issue
for both electronics and new field of condensed matter physics known as
spintronics.\cite{Awschalom,Zutic} One of the problems being actively studied nowadays is
the control on spin (in general, magnetic moment) degrees of freedom for the carriers participating
in the electron transport, optical, magnetization, etc. phenomena. The classical and approved way
to achieve this goal is the application of external magnetic fields which is successfully
used both in fundamental experiments and in commercial device structures. With all its
advantages the application of external magnetic field is not always desirable for
the technological purposes. Hence, alternative methods of spin control are interesting
for both fundamental and applied issues. One of them is the consideration of
the spin-orbit (SO) coupling in those semiconductor nanostructures where it can produce
measurable and potentially usable effects. One type of the SO coupling in heterostructures
is the Rashba coupling\cite{Rash60} coming from the structure inversion asymmetry of confining potential
and effective mass difference. It is important for the experimental purposes that the value of Rashba coupling
strength can be tuned by the external gate voltage\cite{Miller} and reaches the value of $2 \cdot 10^{-11}$ eVm
in InAs-based structures\cite{Grundler} with two-dimensional electron gas (2DEG) which makes its influence
to be quite substantial. Besides, the Dresselhaus term\cite{Dresselhaus} originating due
to the bulk inversion asymmetry is also present in the most commonly used types of heterostructures.
The ratio between the Rashba term with strength $\alpha$ and Dresselhaus term with strength $\beta$ can be
as small as $\alpha / \beta =1.6$ which was reported in the photocurrent experiments.\cite{Ganichev07}
Hence, it seems reasonable to include both Rashba and Dresselhaus terms for more
accurate description of the SO coupling in these structures. The inclusion of both
terms will be further justified if one can find an effect which is sensible to the
particular form of the SO coupling.

In the great variety of spin-dependent properties the problem of field-induced carrier
polarization which accompanies the charge current flow is one of the central ones. Since the pioneer work
by Datta and Das on the concept of spin field-effect transistor\cite{DD} it attracts considerable
attention, and one of the key issues is the spin polarization induced by an external electric field in
the presence of the SO coupling.\cite{Aronov,Edelstein,Raichev} Apart from the great variety of
the results on the challenging problem of spin current, one can mention the calculations of spin
susceptibilities,\cite{esl} the spin polarizations in a bar\cite{Li,Yao,Li2} or in the T-shaped
conductor,\cite{Yamamoto} the interplay between spin and Hall charge current,\cite{Zhao}
the spin accumulation in a quantum wire device,\cite{Wang} the pumping of charge current by spin
dynamics,\cite{Ohe} and the dynamics of localized spins coupled to the conduction
electrons,\cite{Onoda} the injected current-control\cite{Kimura} and detection\cite{Jansen} of spin
accumulation,\cite{kbb,bk} the spin Gunn effect,\cite{Qi} and a recent proposal of the spin
current diode.\cite{Zhai} Another important problem is a possible influence of random nanosize domains of
the SO coupling formed due to the imperfections of the structure \cite{GSh} which can give rise to
the spatially non-uniform character of SO terms leading to the electron spin precession \cite{Liu06}
and the effects referred to the field of spin optics.\cite{KSF}

One of possible ways to create a non-uniform spin distribution in a heterostructure is to apply
a metal-gated superlattice with tunable amplitude of electric potential to the two-dimensional
electron gas (2DEG) with the SO coupling. It was shown by Kleinert, Bryksin and
Bleibaum that in the presence of Rashba SO coupling the external electric field
yields and enhanced spin polarization in a superlattice with a single spin-split band.\cite{kbb}
In addition to the total polarization of the sample, one can be interested in calculating
the local polarization (or spin density) at the point of a real space which may be actually probed
by a detector. It is known that the states with inhomogeneous distribution of spin density
can exist and, which is important, can have long spin relaxation time.\cite{Pershin}
It was shown that such states can be found in a 1D superlattice with Rashba SO coupling
demonstrating a non-uniform distribution of spin density for a given state ${\bf k}$, i.e.,
showing a spin texture.\cite{jetpl} Of course, under the equilibrium conditions the contributions
to the local spin density from all states below Fermi level cancel each other since the populations
of the ${\bf k}$ and $-{\bf k}$ states with the opposite spin projections in a system without magnetic order
are equal. However, in the presence of symmetry breakup during scattering\cite{soscat} or in a non-equilibrium
conditions created by an external radiation\cite{sooptic} one can observe various and controllable spin textures
along the superlattice cell. Hence, it seems also promising to look at the spin polarizations of the charge current
and the the spin textures created by an external DC field applied to the superlattice.

In the present paper we study the DC current, the spin polarization, and the spin textures
in the 1D gated superlattice with simultaneous presence of both Rashba and Dresselhaus SO terms.
The shape of spin textures is calculated as a function of the applied electric field for
fixed values of Rashba and Dresselhaus SO amplitudes $\alpha$ and $\beta$ and also as a function of
$\alpha/\beta$ ratio at fixed electric field and in the interval of $\alpha/\beta$
covering rather wide range of semiconductor materials.
We consider the finite parameters of collision rate and temperature which appear to be non-destructive for
the calculated current and spin characteristics depending on all states below the Fermi level.
The knowledge of field-induced spin textures in addition to the spin polarization and charge current may be
instructive for both fundamental and applied issues of low-dimensional semiconductor structures with strong SO coupling.
It will bee seen that the principal results of the paper regarding the generation of spin textures in a superlattice
can be obtained in a rather elementary model. In this model a simple stationary kinetic equation is considered
with a constant relaxation time and all the calculations of the physical quantities based on the knowledge
of Bloch spinors and the miniband spectrum while keeping in mind that a more general formalism of spin density
matrix can also be applied for more detailed studies.\cite{Aronov,Edelstein,Raichev}

This paper is organized as follows. In Sec.II we briefly describe quantum states and spin polarizations
in the minibands of a superlattice with Rashba and Dresselhaus SO coupling.
In Sec.III we write down the kinetic equation for the distribution function in the presence of
the DC electric field and solve it numerically. The distribution function is used for obtaining the charge
current and the mean spin projections as well as the spin textures. In Sec.IV we calculate and discuss
the spin textures for both fixed Rashba and Dresselhaus amplitudes and for varying $\alpha/\beta$ ratio.
The concluding remarks are given in Sec.V.

\section{Quantum states in SO superlattice}

In this Section we shall briefly describe the quantum states of 2DEG with Rashba and Dresselhaus
SO coupling subject to a one-dimensional (1D) periodic superlattice potential.
A model involving only the Rashba contribution to the SO coupling and a superlattice
potential has been derived previously\cite{jetpl} and applied to the problem of scattering\cite{soscat}
and optical excitation of spin textures.\cite{sooptic}
The Hamiltonian is the sum of the 2DEG kinetic energy operator in a single size quantization band with
effective mass $m$, the Rashba and Dresselhaus SO terms with amplitudes $\alpha$ and $\beta$, respectively,
and the periodic electrostatic potential of the 1D superlattice:

\begin{equation}
\hat{H}=\frac{\hat{p}^2}{2m}+\alpha(\hat{\sigma}_x\hat{p}_y-\hat{\sigma}_y\hat{p}_x)
                            +\beta(\hat{\sigma}_y\hat{p}_y-\hat{\sigma}_x\hat{p}_x)
                            +V(x),
\label{ham}
\end{equation}

where $\hbar =1$ and the periodic potential is chosen in the simplest form
$V(x)=V_0\cos 2\pi x /a$ where $a$ is the superlattice period
and the amplitude $V_0$ can be tuned by the gate voltage.
The eigenstates of Hamiltonian (\ref{ham}) are two-component Bloch spinors with
eigenvalues labeled by the quasimomentum $k_x$ in a one-dimensional Brillouin zone
$-\pi/a \le k_x \le \pi/a$, the momentum component $k_y$, and the miniband index $m$:

\begin{equation}
\psi_{m\bf k}=\sum_{\lambda n} a^m_{\lambda n} ({\bf k})
\frac{e^{i{\bf k}_n{\bf r}}}{\sqrt{2}} \left(
\begin{array}{c}
1 \\
\lambda e^{i\theta_n}
\end{array}
\right),
\quad \lambda = \pm 1.
\label{wf}
\end{equation}

\begin{figure}
  \centering
  \includegraphics[width=85mm]{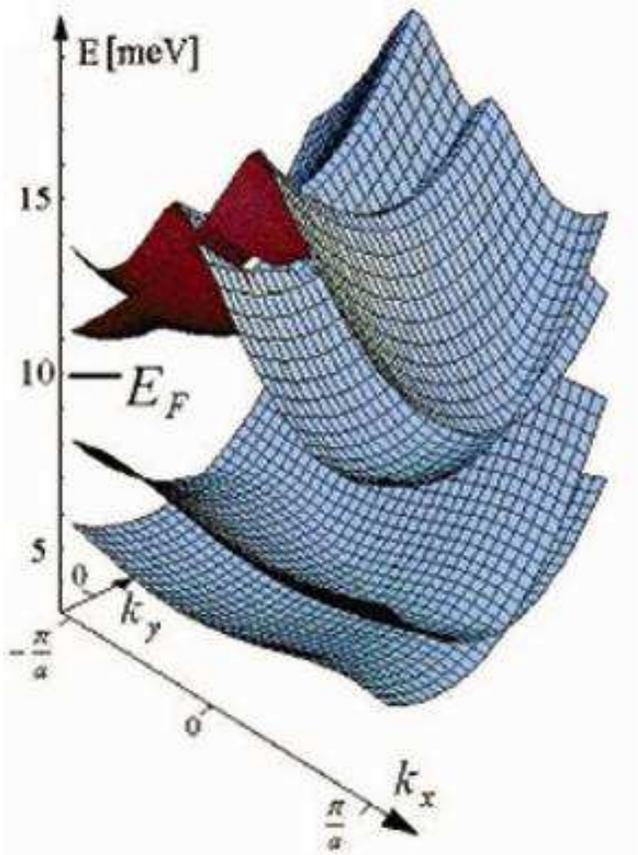}
  \caption{(color online) Energy spectrum of four lowest minibands in the InAs 1D superlattice with
           Rashba SO term $\alpha=2 \cdot 10^{-11}$ eVm and the Dresselhaus term $\beta=1.25 \cdot 10^{-11}$ eVm.
           The other parameters are the electron effective mass $m=0.036$ $m_0$, the superlattice period and
           amplitude $a=60$ nm and $V_0=10$ meV.}
  \label{figband}
\end{figure}

Here ${\bf k}_n={\bf k}+n{\bf b}=\left(k_x+\frac{2\pi}{a}n, \quad k_y\right)$ and
$\theta_n={\rm arg}[\alpha k_y + \beta k_{nx}  - i(\alpha k_{nx}+\beta k_y]$.
The energy spectrum of Hamiltonian (\ref{ham}) consists of pairs of spin-split
minibands. The spacing between two minibands in a pair is mainly determined by the SO parameters
$\alpha$ and $\beta$ while the miniband widths and the gaps in the spectrum are of the order of
the superlattice potential amplitude $V_0$. An example of the energy spectrum is shown in Fig.\ref{figband}
for the four lowest minibands in the InAs-based 1D superlattice with Rashba parameter
$\alpha=2 \cdot 10^{-11}$ eVm plus the Dresselhaus SO term with the amplitude $\beta=1.25 \cdot 10^{-11}$ eVm.
Here the ratio $\alpha / \beta=1.6$ corresponds to the one measured in the photocurrent
experiments on InAs-based structures.\cite{Ganichev07} The cited experiments have shown that the ratio
$\alpha/\beta$ for the most widely used 2D structures varies from $1.5$ for GaAs/AlGaAs and $1.6$ for
InAs/InAlAs quantum wells to $7.6$ in a single GaAs/AlGaAs heterojunction. One can see from Table I
in Ref.\cite{Ganichev07} that none of the samples have shown a negligible impact of the Dresselhaus term
and hence this contribution should be included in the SO part of the Hamiltonian. The other parameters are
the InAs electron effective mass $m=0.036$ $m_0$, the superlattice period $a=60$ nm and the amplitude of
the periodic potential $V_0=10$ meV. It should be noted that the spectrum in Fig.\ref{figband} is limited
to the first Brillouin zone of the superlattice in the $k_x$ direction while the cutoff in the $k_y$ direction
is shown only to keep the limits along $k_x$ and $k_y$ comparable. The minibands in Fig.\ref{figband} have
an inversion symmetry $E({\bf k})=E(-{\bf k})$ but they do not have an additional symmetry with respect
to the change $k_{x,y} \to -k_{x,y}$ if both Rashba and Dresselhaus terms are present.
This feature is very important for the symmetry analysis of the present structure as well as for
the induced spin textures since the absence of the additional symmetry plane
perpendicular to the $y$ axis allows the generation of $S_x$ and $S_z$ components of
spin density for the $x$-oriented electric field, as we shall see below in Sec.IV.
Another argument for the consideration of the lowest available symmetry of the SO term is the possible
influence of random nanosize domains of the SO coupling which can be present due to the imperfections
of the structure.\cite{GSh}

An external electric field changes the occupation distribution in the reciprocal space
and thus it may produce a non-compensated impacts from states with different quasimomenta
to the local spin densities from the symmetrical points in ${\bf k}$-space. Hence, it is instructive
to take a look at the spin polarization described by a vector field
$(\sigma_x({\bf k}),\sigma_y({\bf k}),)$ in the $(k_x,k_y)$ plane. Each of the mean spin projections
is given by

\begin{equation}
\sigma_i({\bf k})=
\langle \psi_{\bf k} \mid {\hat \sigma}_i \mid \psi_{\bf k} \rangle
\label{sk}
\end{equation}

and is calculated for each miniband separately with a given $\psi_{\bf k}$. The spin
vector field is two-dimensional since for both Rashba and Dresselhaus terms the mean
value of $\sigma_z$ is zero. We are interested in the topological structure of
vector field (\ref{sk}) in each miniband since it can provide a justified estimation about
the measurable local spin density which can be induced by an external electric field.

\begin{figure}
  \centering
  \includegraphics[width=85mm]{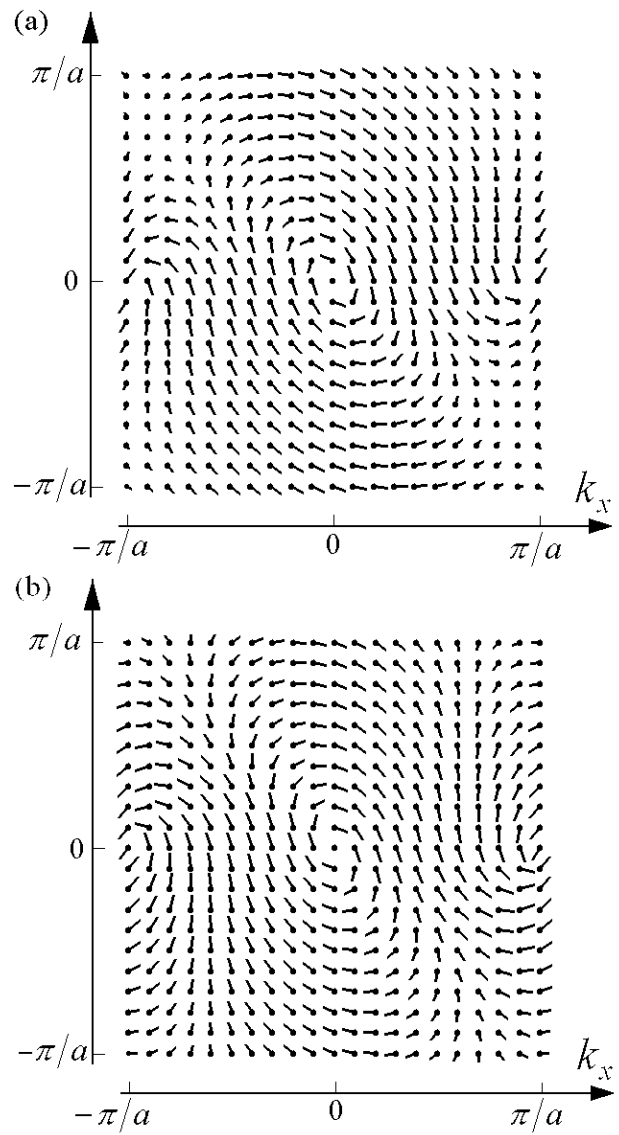}
  \caption{Spin polarization (the origin of vector ${\vec \sigma}({\bf k})$ is marked by black circles)
           (a) in the two lowest and (b) in the next miniband of spectrum in Fig.\ref{figband} with
           the presence of both Rashba and Dresselhaus SO coupling terms with the amplitude
           ratio $\alpha/ \beta = 1.6$.}
  \label{fsk2}
\end{figure}

In Fig.\ref{fsk2} the vector field $(\sigma_x({\bf k}),\sigma_y({\bf k}),)$
is shown schematically for two lowest superlattice minibands with the same parameters
as in Fig.\ref{figband} with both Rashba and Dresselhaus SO terms. In can be seen from Figures \ref{figband}
and \ref{fsk2} that the periodicity of the superlattice along $x$ leads to the same property in
the ${\bf k}$ - space for both energy and spins with the reciprocal lattice vector $(2\pi/a, \, 0)$.
In each miniband the relation

\begin{equation}
{\vec \sigma}({\vec k})=-{\vec \sigma}(-{\bf k})
\label{seq}
\end{equation}

is satisfied, so in equilibrium at each point in the real space the contribution
${\vec \sigma}({\bf k})$ is compensated by the term $-{\vec \sigma}(-{\bf k})$ leaving
the sample non-magnetic. If an external electric field is applied along $x$ direction,
the relation (\ref{seq}) is no longer satisfied, and one can find a non-zero spin accumulation at
the edges of the sample. In the next Sec. we shall calculate this quantity in the presence of
the electric field oriented parallel to the superlattice direction $x$ for Rashba plus Dresselhaus types
of SO coupling.

\section{Charge current and mean spin values}

\subsection{Kinetic equation for the distribution function}

The field-induced distribution of spin density and the charge current can be
calculated with the non-equilibrium stationary distribution function $f_m({\bf k})$
in the miniband $m$ which depends only on the momentum if the stationary and uniform external
electric field $E_x$ is applied along the $x$ direction of the superlattice. We shall neglect
possible non-uniform character of the distribution function in a real space arising due to
the strong non-linearity at high electric field and resulting to the current and the charge
domains of instability. Still, our calculations will include the electric fields high enough
to see the non-linear dependence of both the induced spin densities and the current on
the applied field strength.

In the collision frequency approximation the kinetic equation for $f_m({\bf k})$ under
the conditions described above has a simple form

\begin{equation}
eE_x \frac{\partial f_m({\bf k},E_x)}{\partial k_x}=-\nu [f_m({\bf k},E_x)-F_m({\bf k})],
\label{kineq}
\end{equation}

where $\nu$ is the collision rate and $F_m({\bf k})=1/(1+\exp[(E_m({\bf k})-\mu)/k_B T])$
is the Fermi equilibrium distribution function in the $m$th miniband. In the following we shall assume
that $T=77$ K and $\nu=10^{12}$ $s^{-1}$ which corresponds to the thermal and collision broadening
of $6.6$ and $3.9$ meV, respectively. The position of the Fermi level can be tuned by
the gate voltage and we assume $E_F=10$ meV counted from the bottom of the electron size
quantization band. Such broadening parameters are typical in the experiments
and produce significant smearing of the SO - split miniband structure
shown in Fig.\ref{figband}. However, the charge current, the mean spin values, and the local spin density
are determined by the contributions from all occupied states which makes it survivable under this scale
of broadening. By substituting the dispersion relation $E_m({\bf k})$ obtained in the previous Sec.
for each miniband, the kinetic equation (\ref{kineq}) can be solved directly for the given
strength of the electric field $E_x$, allowing one finding the charge current and the spin density.

\subsection{Mean charge current and spin }

The application of an external electric field generates the charge current through
the superlattice. In the presence of the electric field $E_x$ applied along $x$
the charge current $J_x(E_x)$ and the mean spin values $\sigma_i(E_x)$ can be calculated
directly after obtaining the distribution function from Eq.(\ref{kineq}):

\begin{equation}
J_x(E_x)=e \sum_{m,{\bf k}} \langle \psi_{m{\bf k}} \mid {\hat v}_x \mid \psi_{m{\bf k}}\rangle
                            f_m({\bf k},E_x),
\label{jx}
\end{equation}

\begin{equation}
\sigma_i(E_x)=\sum_{m,{\bf k}} \langle \psi_{m{\bf k}} \mid {\hat \sigma}_i \mid \psi_{m{\bf k}} \rangle
                               f_m({\bf k},E_x),
\label{si}
\end{equation}

\begin{figure}
  \centering
  \includegraphics[width=85mm]{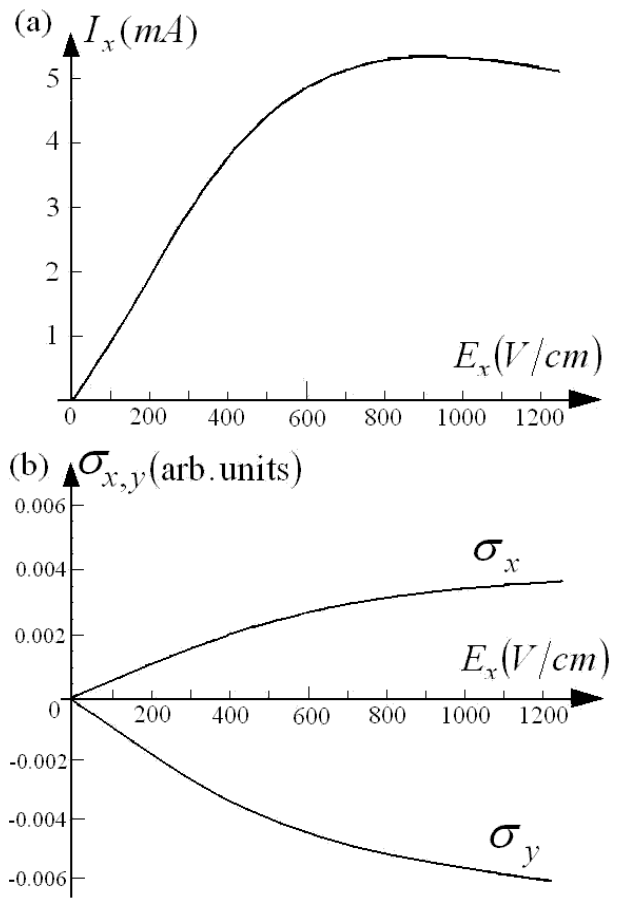}
  \caption{(a) Charge current and (b) mean spin projections induced by the electric field
           in the superlattice with Rashba plus Dresselhaus SO terms. The current in (a) is calculated
           for a $1\, \text{mm} \times 1\text{mm}$ structure with electron density $n=10^{12} \, \text{cm}^{-2}$
           and with all other parameters as in Fig.\ref{figband}.}
  \label{fjs}
\end{figure}

where ${\hat v}_i=\partial {\hat H} / \partial k_i$, and the summation is performed over all
minibands $m$ and all values of $-\pi/a \le k_x \le \pi/a$ and $k_y$, respectively.
The results are shown in Fig.\ref{fjs}(a) for the charge current and in Fig.\ref{fjs}(b) for the mean spin
values. The current in Fig.\ref{fjs}(a) is calculated for a $1 \, \text{mm} \times 1\text{mm}$ structure with
2DEG concentration $n=10^{12} \, \text{cm}^{-2}$ and with all other parameters as for the band
structures in Fig.\ref{figband}. It can be seen from Fig.\ref{fjs}(a) that the low-field Ohmic resistance
of such structure is about $13 \, \text{k}\Omega$. Another well-known feature of the plots in Fig.\ref{fjs}
is the progressive non-linear dependence of the maximum current and spin amplitude which can be seen at high
electric field (greater than $500 \, V/cm$).
Such non-linearity is common for all field-induced quantities in superlattices and is significant when
the Stark frequency $\Omega=|e|E_x a/\hbar$ becomes comparable with the collision frequency $\nu$.

It is known that in the presence of Rashba SO term the only non-vanishing component of
the $E_x$ electric field-induced accumulated spin is $\sigma_y$.\cite{kbb} In the presence of both Rashba and
Dresselhaus terms the $\sigma_x$ component can also be nonzero, as it is evident by looking onto
the topology of the spin vector field in Fig.\ref{fsk2}. This topological aspect is confirmed by the calculations of
mean spin value dependencies on the electric field which are shown in Fig.\ref{fjs}(b). One can see that $\sigma_x$
and $\sigma_y$ can have an equal magnitude as long as the Dresselhaus term is comparable to the Rashba term.
Thus, the consideration of both Rashba and Dresselhaus terms seems to be important for the calculation of
non-vanishing spin components measured experimentally. Another spin-related quantity that we shall discuss below and
which is actually measured in the experiments is the local spin density in a real space determined by all states below
the Fermi level. We shall see that this spin densities can be spatially non-uniform in the presence of
the superlattice potential.

\section{Spin textures}

Since the SO coupling is present, the spin polarization of charge carriers may also take place,
resulting in a spin accumulation in a superlattice.\cite{kbb} The spin-related quantity which
can actually be probed by a tip in the experiment or possibly utilized in a spintronic
nanostructure device is the local spin density in a real space, which manipulation is one
of the primary goals of spintronics. The local change of the spin density is often referred
to the conception of a spin current which permanently attracts a considerable attention
of researchers\cite{esl,bk,Sherman,Shi,Sugimoto} (only few papers from a great number of published articles
on the spin current are cited here as an example), and where the different definitions have been proposed.
Taking into consideration the goals of the present paper, one should mention that the experimental
studies of the spin current phenomena are presently focused on the observation of spin
accumulation. Indeed, one can observe an equal change of local spin density caused by two different
processes: the first one is the transport of spin-polarized charge carriers while the second one
is the local "rotation" of spins which is not always accompanied by the carrier transfer.
In both cases the actually measured observable is the local spin density which change, as we see,
does not strictly require involving the conception of spin current.
The problem of the spin current definition remains to be one of the hottest topics on spintronics
since the work of Rashba\cite{Rashba} where the possibility of the existence of spin currents even
in the equilibrium has been demonstrated. The different possibilities of local spin evolution are reflected
also in the non-conservation of the spin in terms of the continuity equation
$\partial S_i/ \partial t + (\nabla \cdot {\bf J}_i)=T_i$. If the spin current density is defined
as $J_i^{j}=\text{Re} \left[ \psi^{\dagger} \frac{1}{2}\{ v_j,\sigma_i \} \psi \right]$,
than one has to introduce the torque density
$T_i=\text{Re} \left[ \psi^{\dagger} \frac{1}{i}[\sigma_i,H ] \psi \right]$
in the right side of the continuity equation. It was shown that this equation takes the usual form with zero
in the right side if another definition $J_i^{j}=\frac{d}{dt}(x^j \sigma_i)$ of spin current is considered.\cite{Shi}
The proper and uniform definition of spin current based on the measurable quantities is still under discussion,
and the goal of the present paper is the local measurable spin density rather than the spin current.
Hence, below we shall focus on the electric field-induced local spin density which is non-uniform in
the superlattice cell and thus can be referred as the spin texture. It should be stressed
that this texture accompanies the charge current and thus provides the information about the tunable spin
polarization of the electrical current which is of big importance for possible device applications.

\subsection{Fixed $\alpha/\beta$ and variable electric field}

The Rashba and Dresselhaus SO terms in a Hamiltonian of the free particle produce a spin polarization for
the plane wave spinor $\psi_{\bf k}$ with a given ${\bf k}$ which has a uniform spin density distribution
in the real space. If an additional superlattice potential is applied, the spin density for $\psi_{m {\bf k}}$
in the $m$th miniband becomes non-uniform and forms the spin texture.\cite{jetpl} In the equilibrium conditions
the spin densities from all states below $E_F$ cancel each other which leads to zero spin
density in any point of the real space. If the equilibrium is destroyed by an external
electric field, one can expect to measure not only the non-zero spin
accumulation\cite{kbb} but also the local spin density which varies along the
superlattice cell and forms variable spin textures which shape can be modified by manipulating
the system parameters.

\begin{figure}
  \centering
  \includegraphics[width=80mm]{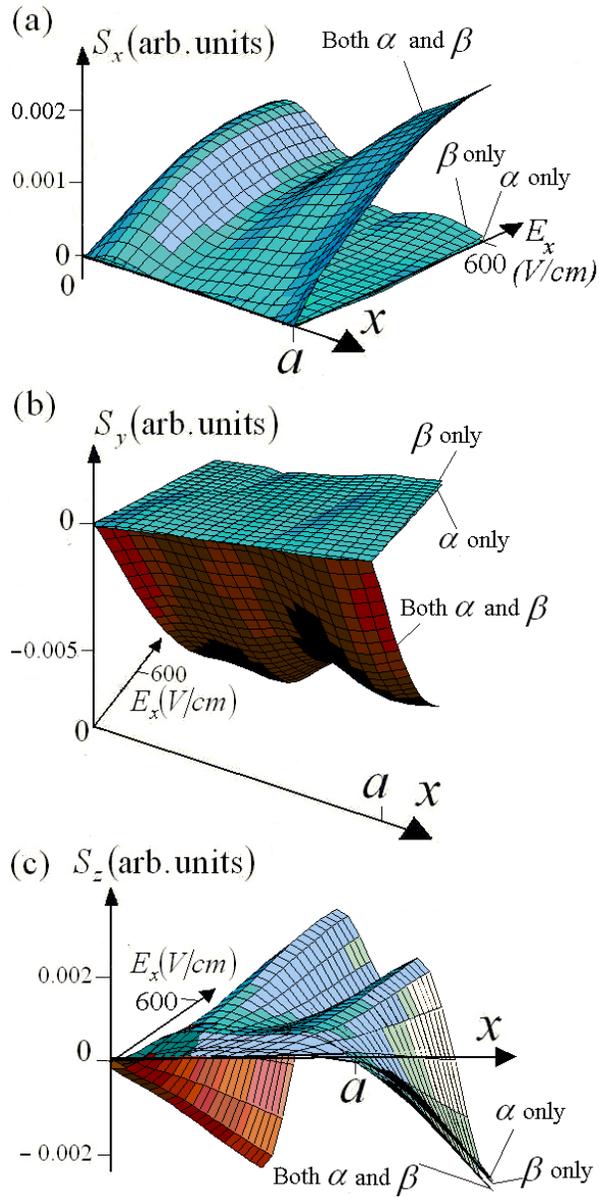}
  \caption{(color online) Electric field dependence of spin texture (a) $x$-component (b) $y$ -component and
           (c) $z$ - component shown in one superlattice cell $0 \le x \le a$ for the field interval
           $0 \le E_x \le E_{\text{max}}$. The textures for two limiting cases with $\beta=0$ and $\alpha=0$ as well
           as for the fixed Rashba/Dresselhaus ratio $\alpha/\beta=1.6$ corresponding to InAs-based structure are shown.
           The temperature $T=77$ K, the collision rate $\nu=10^{12}$ $s^{-1}$, and all other parameters are the same
           as in Fig.\ref{figband}.}
  \label{frd}
\end{figure}

\begin{figure}
  \centering
  \includegraphics[width=80mm]{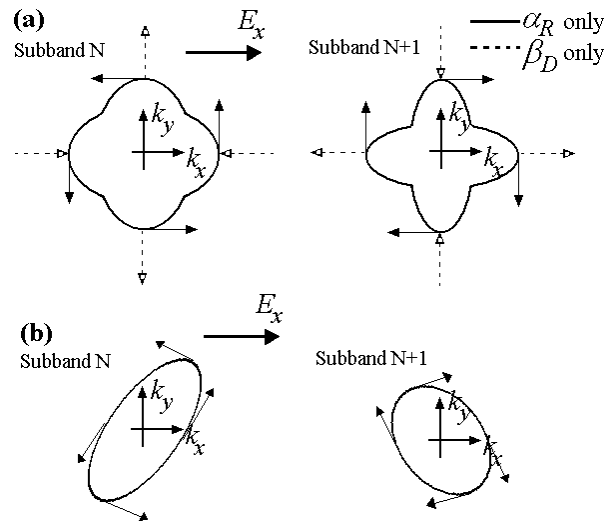}
  \caption{(a) Schematic view of mean spin alignment for two SO-split subbands for pure Rashba (solid arrows)
           and pure Dresselhaus (dashed arrows) SO coupling which energy spectra are symmetrical with respect
           to both $k_x \to -k_x$ and $k_y \to -k_y$ transformations, leading to a high degree of cancellation of
           the spin textures from the neighboring subbands induced by the electric field $E_x$.
           (b) The same for the presence of both Rashba and Dresselhaus terms where the spectra are invariant
           only with respect to the $(k_x,k_y) \to (-k_x,-k_y)$ transformation, leading to the increasing differences
           in the energy dispersion shapes and creating a much smaller degree of cancellation from the neighboring
           subbands, thus increasing the spin texture amplitude (see Fig.\ref{frd}).}
  \label{fexplain}
\end{figure}

In this subsection we are interested in calculating the spin density $S_i(x,E_x)$ along the superlattice cell
as a function of the electric field strength $E_x$. If the electric field is applied along $x$ than
the density depends only on $x$ in the real space since in the $y$ direction the system is totally
homogeneous. After obtaining the distribution function from Eq.(\ref{kineq}), one can write

\begin{equation}
S_i(x,E_x)=\sum_{m,{\bf k}}\left( \psi^{\dagger}_{m{\bf k}} {\hat \sigma}_i \psi_{m{\bf k}} \right)
                   f_m({\bf k},E_x),
\label{sr}
\end{equation}

where the summation and integration are performed over all minibands $m$ and all values
of $-\pi/a \le k_x \le \pi/a$ and $k_y$, respectively. The results can be presented in
a form of a 3D plot showing each of the spin density components $S_i$ separately as a
function of the position $x$ inside the superlattice cell and the electric field
strength $E_x$.

In Fig.\ref{frd} we show the field-induced spin textures in a 2DEG for two limiting cases with
$\beta=0$ and $\alpha=0$ as well as for the fixed Rashba/Dresselhaus ratio $\alpha/\beta=1.6$,
taking all other system parameters as in Fig.\ref{figband}. The plots in Fig.\ref{frd} and below
in Fig.\ref{fratio} show each component of spin ($S_x(x)$ in part (a), $S_y(x)$ in part (b) and $S_z(x)$ in part (c))
separately on the vertical axis calculated as a $z=f(x,y)$ function of the position $x$ in a superlattice cell
($x$ axis) and of the applied electric field ($E_x$ axis in Fig.\ref{frd}) or of the $\alpha/\beta$ ratio
($\alpha/\beta$ axis in Fig.\ref{fratio}). At zero electric field the structure is maintained in the thermodynamic
equilibrium with equal population of the ${\bf k}$ and $-{\bf k}$ states in the reciprocal space having
the opposite spin projections. Thus, without the electric field-induced imbalance of this
population the structure is not expected to demonstrate non-zero spins in any point of the real space,
i.e. the spin textures in our system have a non-equilibrium origin.

The most striking feature of spin textures in Fig.\ref{frd} is the big amplitude difference of
the field-induced spin texture components $S_x$ and $S_y$ for limiting cases $\beta=0$ and $\alpha=0$ and for
a general case $\alpha/\beta=1.6$ while the $S_z$ amplitude is rather unaffected by the $\alpha/\beta$ variations.
The explanation of this effect is coming from the symmetry considerations as well as from
the analysis of the SO superlattice subband energy spectrum which is presented schematically in Fig.\ref{fexplain}.
First of all, it should be mentioned that in the superlattice with SO coupling the
energy subbands always come in pairs. Inside each pair the dispersion surfaces are relatively weakly split by the SO
coupling (see Fig.\ref{figband}) and their contributions to the induced spin textures are in generally of the opposite
sign (see Fig.\ref{fsk2}). In order to illustrate this we plot in Fig.\ref{fexplain}(a) a very simple schematic view
of mean spin alignment in the ${\bf k}$-space for two SO-split superlattice subbands for pure Rashba (solid arrows)
and pure Dresselhaus (dashed arrows) SO coupling which energy spectra are symmetrical with respect to both $k_x \to -k_x$
and $k_y \to -k_y$ transformations, leading to a high degree of cancellation of the field-induced spin textures from
the neighboring subbands. In Fig.\ref{fexplain}(b) the same view is shown for the presence of both
Rashba and Dresselhaus terms where the spectra are invariant only with respect to the $(k_x,k_y) \to (-k_x,-k_y)$
transformation, leading to the increasing differences in the energy dispersion shapes and creating a much smaller degree
of cancellation from the neighboring subbands, thus increasing the spin texture amplitude (see Fig.\ref{frd}).
The treatment of $S_z$ component of spin textures cannot be handled in the same way since its mean value for a quantum
state of our Hamiltonian is zero, meaning that the condition

\begin{equation}
\int_{0}^a S_z(x)dx=0
\label{szero}
\end{equation}

is always fulfilled. The local non-zero $S_z(x)$ component is formed by the effective
magnetic field $\propto [\nabla V(x), {\vec p}]$ arising due to the superlattice potential
$V(x)$. Since this potential is periodic, the mean value of its gradient is zero which
is reflected in Eq.(\ref{szero}). These general properties of the local non-zero $S_z(x)$ component are related only
to the spatial dependence of the superlattice potential and on the applied electric field and thus they should not depend
strongly on the precise value of Rashba/Dresselhaus ratio as long as the SO coupling is present.
This expectation is consistent with the plots for $S_z(x)$ in Fig.\ref{frd}(c)
where the textures for three different sets of parameters are very close to each other.
Since the local spin density is an actually measurable quantity in the experiments, the creation of field-induced spin
textures in superlattices with both Rashba and Dresselhaus SO terms should be taken into consideration for possible
experimental and device purposes.

\subsection{Variable $\alpha/\beta$ and fixed electric field}

\begin{figure}
  \centering
  \includegraphics[width=75mm]{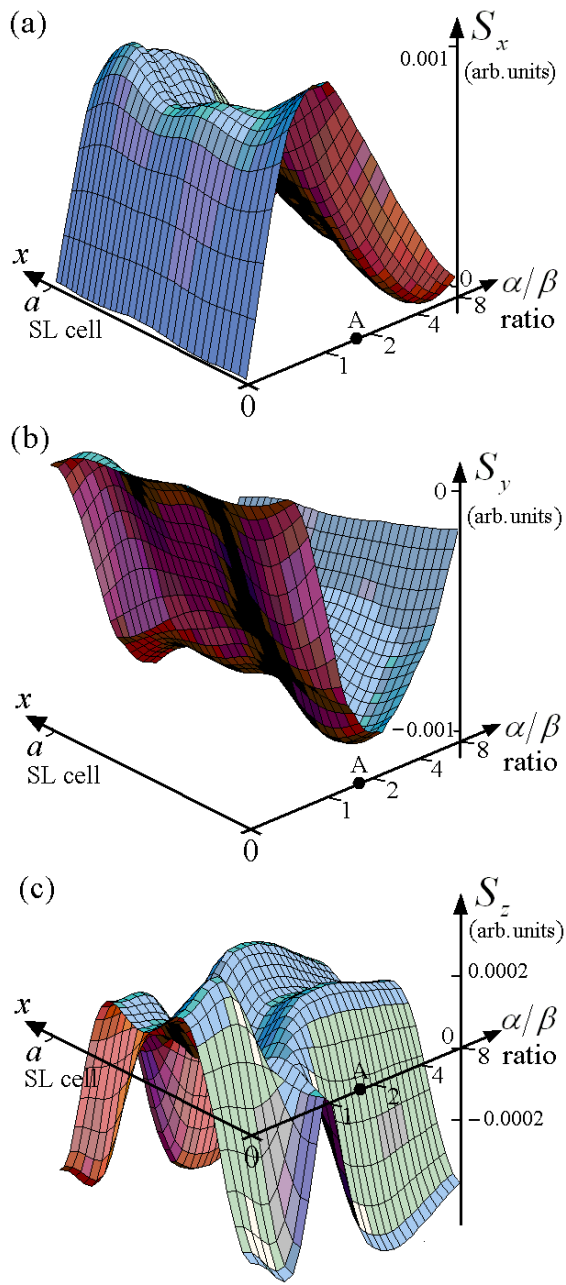}
  \caption{(color online) The dependence on Rashba - to - Dresselhaus parameter ratio $\alpha/\beta$
           of the spin textures in the superlattice cell $0 \le x \le a$ for (a) $x$-component
           (b) $y$ -component and (c) $z$ - component induced at fixed electric field
           $E_x = 50$ V/cm. The point "A" corresponds to the example of InAs-based structure with
           $\alpha/\beta=1.6$ considered above. All other parameters are the same as in Fig.\ref{frd} and
           Fig.\ref{figband}. The electric field-induced spin textures have the biggest amplitudes in
           a region of comparable Rashba and Dresselhaus contributions to spin-orbit coupling.}
  \label{fratio}
\end{figure}

Now we shall consider the dependence of spin texture components on the ratio $\alpha/\beta$ of Rashba and
Dresselhaus contributions to the SO coupling when the electric field strength $E_x=50$ V/cm is fixed in a low-field
regime of the interval in Figures \ref{fjs},\ref{frd} which is more desirable for practical purposes.
The upper limit of $\alpha/\beta$ can be estimated from the actual photocurrent experiments data\cite{Ganichev07}
where in different structures constructed from different materials this ratio has been reported to vary from $1.5$ to $7.6$.
Taking this into consideration, we shall restrict ourselves to the interval $0 \le \alpha/\beta \le 8.0$.
Keeping all other parameters of the system unchanged, we obtain the $\alpha/\beta$ - dependencies of spin texture components
in a superlattice cell which are shown in Fig.\ref{fratio}. The point "A" on the $\alpha/\beta$ axis corresponds
to the example of InAs-based structure with $\alpha/\beta=1.6$ considered in the previous parts of the paper, and
the position of the vertical axis is displaced for a better view.

First, let us examine the lower part of the interval when $\alpha/\beta <1$, i.e. when the SO coupling is dominated by
the Dresselhaus term in a macroscopically symmetrical semiconductor structure with significant bulk inversion
asymmetry (BIA) and negligible structure inversion asymmetry (SIA). One can see from Fig.\ref{fratio} that the only
significant component of spin texture in this limit is $S_z(x)$ although it is always several times smaller in amplitude
than the maximum achievable $S_x(x)$ and $S_y(x)$ components at various $\alpha/\beta$. This local component of
field-induced spin texture can be non-zero since the structure has a microscopic BIA and is non-homogeneous due
to the presence of the superlattice. Nevertheless, the mean value of the out-of-plane component $S_z(x)$ is zero
for all values of $\alpha/\beta$ and $E_x$, i.e. the condition (\ref{szero})
is always fulfilled, as it can be checked numerically for the textures in Fig.\ref{frd}(c) and Fig.\ref{fratio}(c).

Another important issue is the symmetry properties of the induced polarization. The presence of purely Dresselhaus
SO coupling brakes the bulk inversion symmetry, i.e. the ${\bf r} \to -{\bf r}$ element of symmetry is
no longer present. For the induced spin texture considered here the only component of spin is $S_z$, i.e.
${\bf S}=(0,0,S_z)$. Despite the existence of the $C_2$ rotation axis parallel to the $x$ direction,
the application of this rotation to the particular orientation of the induced spin ${\bf S}(E(x))=(0,0,S_z(E_x))$
is equivalent to the inversion ${\bf r} \to -{\bf r}$ which is no longer present as a symmetry element due
to the Dresselhaus SO coupling. As for the pure Rashba coupling, the same effect is produced by the breaking of
the $z \to -z$ symmetry by the SIA with the confinement potential $U(z) \ne U(-z)$.
Hence, the existence of the electric field-induced $S_z$ spin component here is consistent
with the symmetry relations. It should be mentioned also that the existence of the out-of-plane spin polarization
induced by the in-plane electric field in a system with both Rashba and Dresselhaus SO coupling is a well-known phenomenon
in the context of the spin Hall effect where the uniformly spaced spin currents of $S_z$ components have been
predicted.\cite{Shen,Sinitsyn,LiShen,Chen,Xing}

We now consider the general case of non-zero Rashba and Dresselhaus SO terms.
As we move along the $\alpha/\beta$ axis, it becomes clear that the maximum amplitudes for $S_x(x)$ and $S_y(x)$
components are achieved when Rashba and Dresselhaus terms become comparable in size while the shape of $S_z(x)$
is only slightly modified. The $S_x(x)$ and $S_y(x)$ components both have a non-zero
mean value which is in the agreement of the calculated spin projections in Fig.\ref{fjs}
and is important in the scope of the practical issues of spin accumulation. The presence of both Rashba and
Dresselhaus terms leads to the appearance of two non-zero components of the accumulated spin instead of one ($S_y$ only)
when the SO coupling is purely of Rashba type.\cite{kbb}
It should be noted that a high degree of spin polarization at equal strengths of Rashba and Dresselhaus terms is
in the agreement with the results of numerous studies of transport and spin Hall phenomena in such systems.
Hence, the example of $\alpha/\beta=1.6$ for InAs-based structure which was studied above in
details is promising since the desirable spin properties in such system are manifested on the highest achievable level.

Finally, let us consider the limit $\alpha \gg \beta$ when the Rashba term dominates over the Dresselhaus
term, although the latter is reported still to be non-zero in practically used structures,\cite{Ganichev07} and thus
the symmetry of the system on the whole $\alpha/\beta$ axis remains to be the same (except for one point $\alpha/\beta=0$).
Here one can observe a non-zero integral over $S_y(x)$ only which is consistent with previous results on the $S_y$
component of spin accumulation in a system with pure Rashba SO coupling.\cite{kbb} The $S_x(x)$ component tends to vanish
while the $S_z(x)$ component has qualitatively the same form on the whole $\alpha/\beta$ axis and zero mean value.
We see that the shape and the amplitude of $S_z(x)$ are only weakly dependent on
the specific value of $\alpha/\beta$, as it has been discussed in the previous subsection.
One can see a change of the $S_z(x)$ shape occurring at the transmission trough the point $\alpha=\beta$ where
the function $S_z(x)$ almost vanishes which is in agreement with topological properties of the spin structure.
However, the major part of experimentally studied nanostructures with SO coupling are characterized
by some intermediate ratio $\alpha/\beta$ which is far away from any of the special value.\cite{Ganichev07}
We believe that further theoretical and experimental studies of gated spin-orbit structures are promising since
the strength of Rashba term can be widely tuned by the gate voltage,\cite{Miller} and the generation of spin
textures presented in Figures \ref{frd} and \ref{fratio} with different ratios $\alpha/\beta$ including the special ones
seems to be experimentally accessible.

\section{Conclusions}

We have studied the DC charge current, the spin polarization, and the spin textures in the 1D gated
superlattice with both fixed and varying Rashba and Dresselhaus SO coupling terms and the spectrum consisting of
multiple pairs of spin-split minibands. We have seen how the presence of both Dresselhaus and Rashba terms with varying
ratio is reflected in the SO - sensitive spin properties of the electron gas for both mean spin values and spin textures.
It was found that the spin component with zero mean value can have non-vanishing field-induced spin
texture in a superlattice cell which can be probed experimentally. It is shown that the consideration
of the finite parameters for collision rate and temperature is non-destructive for the calculated
current and spin characteristics depending on all states below the Fermi level.
The knowledge of field-induced spin textures in addition to the spin polarization and charge current
may be instructive for both fundamental and applied issues of low-dimensional semiconductor structures
with strong SO coupling.

\section*{Acknowledgments}

The author is grateful to V.Ya. Demikhovskii, E.Ya. Sherman and A.A. Perov for helpful discussions.
The work was supported by the RNP Program of the Ministry of Education and Science RF,
by the RFBR, CRDF, and by the Foundation "Dynasty" - ICFPM.

\end{document}